\documentclass[pra,aps,amsmath,amssymb,amsfonts,twocolumn,nofootinbib,longbibliography,superscriptaddress]{revtex4-2}
\usepackage{amssymb}
\usepackage{bm,mathrsfs}
\usepackage{graphicx}
\usepackage{subfigure}
\usepackage{epsfig}
\usepackage{amsmath,bbm}
\usepackage{amsfonts,amssymb}
\usepackage{times}
\usepackage{verbatim}
\usepackage[sort&compress]{natbib}
\usepackage{amsmath}
\usepackage[colorlinks=true,citecolor=blue,linkcolor=blue,urlcolor=blue]{hyperref}
\usepackage[usenames]{color}
\usepackage[section]{placeins}
\usepackage{multirow}
\usepackage{booktabs}

\newcommand{\bra}[1]{\langle #1|}
\newcommand{\ket}[1]{|#1\rangle}

\begin{document}
%%%%%%%%%%%%%%%%%%%%%%%%%%%%%%%%%%%%%%%%%%%%%%%%%%%%%%%%%%%%%%%%%%%

\title{Quantum battery based on dipole-dipole interaction and external driving field}

\author{Wuji Zhang}
\affiliation{Center for Quantum Sciences and School of Physics, Northeast Normal University, Changchun 130024, China}

\author{Shuyue Wang}
\affiliation{Center for Quantum Sciences and School of Physics, Northeast Normal University, Changchun 130024, China}

\author{Chunfeng Wu}
\affiliation{Science, Mathematics and Technology, Singapore University of Technology and Design, 8 Somapah Road, Singapore 487372, Singapore}

\author{Gangcheng Wang}
\email{wanggc887@nenu.edu.cn}
\affiliation{Center for Quantum Sciences and School of Physics, Northeast Normal University, Changchun 130024, China}

\date{\today}

\begin{abstract}
The Dicke model is a fundamental model in quantum optics, which describes the interaction between quantum cavity field and a large ensemble of two-level atoms. In this work, we propose an efficient charging quantum battery achieved by considering an extension Dicke model with dipole-dipole interaction and an external driving field. We focus on the influence of the atomic interaction and the driving field on the performance of the quantum battery during the charging process and find that the maximum stored energy exhibits a critical phenomenon. The maximum stored energy and maximum charging power are investigated by varying the number of atoms. When the coupling between atoms and cavity is not very strong, compared to the Dicke quantum battery, such quantum battery can achieve more stable and faster charging. In addition, the maximum charging power approximately satisfies a superlinear scaling relation $P_{\rm max}\varpropto\beta N^{\alpha}$, where the quantum advantage $\alpha=1.6$ can be reached via optimizing the parameters.
\end{abstract}

\maketitle
\section{Introduction}
\label{SecI}
With the advancement of quantum technology, there is a growing interest in schemes that utilize quantum effects to enable superior performance of future technological devices \cite{PhysRevA.102.012822,PhysRevB.104.014519,PhysRevA.100.022113,RevModPhys.92.015003,PhysRevX.2.031007,PhysRevA.97.013851}. Recently, it has shown great success in several practical fields, such as quantum computing, quantum cryptography, and thermodynamic nanoscale device, which are expected to completely solve data analysis in the communication process, optimize sensitive parameters to improve network security, and provide more accurate temperature measurement \cite{PhysRevLett.101.040501,PhysRevLett.89.057902,RevModPhys.74.145,PhysRevLett.103.230501,PhysRevE.103.052104,PhysRevB.97.155404}. Overall, the development of quantum technology promises to offer more miniaturized and more precise devices. The devices with potential quantum information processing have also been developed, but strategies for storing and releasing energy in these devices remain a major problem to be addressed \cite{PhysRevA.97.022311,PhysRevA.62.042305,PhysRevA.67.062316}. 

The question of whether quantum effects can improve energy storage to meet the current needs of quantum devices has been explored. The so-called quantum battery (QB), a small quantum system for energy storage and extraction, has been published in a seminal paper by Alicki and Fannes \cite{PhysRevE.87.042123}. This provides an efficient way to solve the energy constraints of quantum devices \cite{PhysRevResearch.2.023095,PhysRevLett.122.210601,PhysRevE.101.062114}. A central goal for such research field is to optimize the performance of QBs, such as energy storage and charging rate \cite{PhysRevA.100.043833,PhysRevLett.128.140501,PhysRevResearch.2.043196,PhysRevE.106.014138}. Generally speaking, two distinct schemes for charging have been proposed, namely collective charging and parallel charging \cite{Binder_2015,PhysRevLett.118.150601}. For collective charging scheme, all cells are charged from the same charger; for parallel charging, each cell is charged through its own charger independently. Some results have demonstrated that collective charging shows better performance over parallel charging and charging acceleration can be achieved \cite{PhysRevA.97.022106,PhysRevLett.125.236402,PhysRevE.105.064119,PhysRevResearch.4.043150,PhysRevE.104.024129}. The advantage of collective charging over parallel charging is called the quantum charging advantage \cite{PhysRevLett.122.047702,PhysRevE.99.052106}. 

 In pursuit of the quantum advantage and possible experimental realization, the QB has been proposed in various models, such as two-level systems \cite{PhysRevE.103.042118,PhysRevLett.129.130602}, three-level systems \cite{Dou2021,PhysRevE.100.032107,Dou_2020}, two photons model 
\cite{PhysRevB.102.245407}, the superconducting circuit model 
\cite{Hu_2022,Zheng_2022,PhysRevA.107.023725}, Lipkin-Meshkov-Glick model 
\cite{PhysRevResearch.2.023113}, Sachdev-Ye-Kitaev model 
\cite{PhysRevLett.125.236402,PhysRevA.105.L010201,Rosa2020}, Heisenberg spin-chain model
\cite{PhysRevA.101.032115,PhysRevE.102.052109}, quantum cavity model \cite{PhysRevB.105.115405,PhysRevA.106.032212}, collision model \cite{PhysRevE.106.054119,morrone2022charging,PhysRevResearch.5.013155}, many-body localized model \cite{PhysRevE.105.044125,PhysRevB.99.205437,PhysRevA.104.L030402,PhysRevB.100.115142}, dissipation model \cite{PhysRevE.104.064143,PhysRevA.105.062203} and so on \cite{rodriguez2022optimal,rodriguez2023catalysis}. One of the most well-known examples is the Dicke model \cite{PhysRevA.103.033715,PhysRevResearch.2.023113,PhysRev.93.99}, which describes the interaction of the ensemble of two-level atoms with the single-photon mode of a cavity. Despite the relative simplicity of the Dicke model, it still exhibits several interesting phenomena as the coupling strength increases to ultrastrong coupling (USC) or even deep-strong coupling (DSC) \cite{PhysRevLett.117.210503,PhysRevA.96.013849,PhysRevA.102.013701}. Recently, QBs based on the Dicke model have been introduced, assuming a conventional coupling between atoms and photon radiation of a cavity. It has been reported that a $\sqrt{N}$ acceleration can be achieved for Dicke QB under the collective charging scheme \cite{PhysRevLett.103.083601,PhysRevLett.120.117702}.

In the development stage of QBs, the spin-chain model placed in a optical cavity has also received much attention \cite{PhysRevResearch.4.013172,PhysRevA.105.022628,PhysRevA.105.022628}. The cavity can affect the radiation of the atoms, inducing atom-atom interactions and collective properties. It is worth noting that various protocols can be experimentally employed to create an array of traps in the cavity to maintain the atoms at a fixed distance \cite{Mishina_2014,PhysRevLett.99.213601}. The atoms placed at a smaller distance are coupled to each other by dipole-dipole interaction \cite{PhysRevLett.114.113002,PhysRevA.82.033412,PhysRevA.106.013703}. Previous studies have rarely considered long-distance interactions of atoms, but long-distance interactions between atoms can produce many interesting phenomena that cannot be ignored in actual atomic chain models \cite{PhysRevA.81.022303,PhysRevA.70.052302}. These interactions are dependent on the distance between atoms. Through the interaction, we can regulate the energy levels and control whether atomic transitions occur \cite{PhysRevA.82.033412,PhysRevA.94.013420,PhysRevLett.119.053202,PhysRevA.95.023807,PhysRevB.105.205416}. A natural question arises: can the charging process be accelerated compared to Dicke QB when considering long-distance interactions between atoms?

On the other hand, in a recent work \cite{PhysRevB.105.115405}, a generalized Dicke QB has been proposed based on the global entanglement interactions and the driving field, which can achieve a much faster charging process than Dicke QB. To take another step forward, we consider the direct dipole-dipole interaction that depends on the atomic distance. In this study, we present the atoms with dipole-dipole interaction that are placed in the driven optical cavity as a QB system. We discuss the essential factors of the QB such as stored energy, charging power and energy quantum fluctuation under various conditions. In addition, we also introduce the quantum advantage of maximum charging power to analyze the performance of QBs under the collective charging scheme.
 
The paper is organized as follows. In Sec. \ref{SecII}, we propose the QB model, charging protocols, and numerical approach. In Sec. \ref{SecIII}, we analyze the influence of the external driving field and dipole-dipole interaction between atoms on the performance at different coupling parameters. In Sec. \ref{SecIV}, the impact of the number of qubits in the model on the maximum stored energy and maximum charging power under various coupling conditions and the quantum advantage of maximum charging power are analyzed. Finally, we give a summary that our QB can be more efficient than Dicke QB in Sec. \ref{SecV}.

\section{Model of the battery and approach of charge}
\label{SecII}
\begin{figure}[htbp]\centering
\includegraphics[scale=0.3]{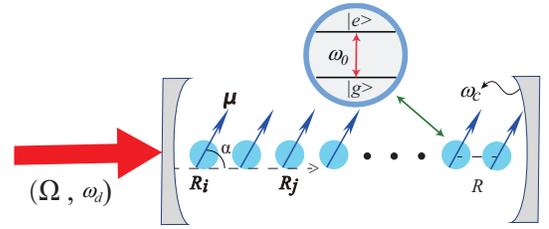}
\caption{Schematic diagram of $N$ qubits coupled to each other by dipole-dipole interaction. The unique cavity with frequency $\omega_{c}$ is driven by a coherent laser field with strength $\Omega$ and frequency $\omega_{d}$. $R$ denotes the distance between adjacent atoms. $\alpha$ denotes the angle between the dipole moment ($\bm{\mu}$) and atomic position vector ($\bm{R_{i}}$). Each atom can be considered as a two-level system with the ground state ($\ket{\textit{g}}$) and excited state ($\ket{\textit{e}}$).}
\label{Fig1}
\end{figure}

The proposed QB consists of $N$ identical two-level atoms coupled to a single-mode optical cavity with an external driving field. Here we consider that two-level atoms (qubits) are placed at a small distance, leading to the presence of dipole-dipole interaction \cite{PhysRevA.105.052439}. The Hamiltonian of this QB system reads (we set $\hbar=1$ throughout this work)
\begin{equation}\label{eq_01}
  \hat{H}_{s}(t) = \hat{H}_{b} + \Theta(t)\hat{H}_{c},
 \end{equation}
where $\Theta(t)=\theta(t)-\theta(t-T)$ with $\theta(t)$ being the Heaviside step function. Such function represents a sudden switch on/off of the charging process: $\Theta(t)=1$ for $0\leq t \leq T$ with $T$ being the charging interval; $\Theta(t)=0$ elsewhere. The terms $\hat{H}_{b}$ and $\hat{H}_{c}$, describing the battery and charger, can be written as
\begin{equation}\label{eq_02}
\begin{split}
   \hat{H}_{b}=\omega_{0}&\hat{J}_{z}, \\
\hat{H}_{c}=\omega_{c}\hat{a}^{\dagger}\hat{a}+2g(\hat{a}^{\dagger}+\hat{a})&\hat{J}_{x}+\hat{H}_{\rm int}+\hat{H}_{d},\\
  \end{split}
\end{equation}
where
\begin{equation}\label{eq_03}
\begin{split}
   \hat{H}_{\rm int}&=\sum_{i\neq j}^{N}\eta_{ij}(\hat{\sigma}_{i}^{-}\hat\sigma_{j}^{+}+\hat\sigma_{j}^{-}\hat\sigma_{i}^{+}),\\
   \hat{H}_{d}&=\Omega\cos(\omega_{d}t)(\hat{a}^{\dagger}+\hat{a}).
  \end{split}
\end{equation}
The notation $\omega_{0}$ denotes the energy splitting between the atomic ground state $\ket{\textit{g}}$ and atomic excited state $\ket{\textit{e}}$, $\hat{J}_{\alpha} = \frac{1}{2}\sum_{i=1}^{N} \hat{\sigma}_{i}^{\alpha}$ ($\alpha=x,y,z$) are the collective spin operators with $\hat{\sigma}_{i}^{\alpha}$ being the Pauli operators acting on the $i$-th site. The operators $\hat{J}_{\alpha}$ constitute a representation of the $\mathfrak{su}(2)$ algebra, obeying the usual commutation relations: $[\hat{J}_{\alpha}, \hat{J}_{\beta}] = i\varepsilon_{\alpha\beta\gamma} \hat{J}_{\gamma}$ ($\alpha, \beta, \gamma \in\{x, y, z\}$) with $\varepsilon_{\alpha\beta\gamma}$ being the Levi-Civita symbol. The notation $\hat{a}$ ($\hat{a}^{\dagger}$) is the annihilation (creation) operator of the cavity field with frequency $\omega_{c}$. Otherwise, $g$ denotes the uniform atom-cavity coupling strength. The cavity is pumped by a coherent laser with amplitude $\Omega$ and frequency $\omega_{d}$. The symbol $\eta_{ij}$ denotes the dipole-dipole interaction between the $i$-th and the $j$-th atoms, which is given by \cite{PhysRevA.105.052439,PhysRevA.89.043838,PhysRevA.70.022511}
\begin{equation}\label{eq_04}
\eta_{ij}=-\frac{3}{4}\frac{\Gamma_{0}c^{3}}{\omega_{0}^{3}|(i-j)R|^{3}}\left(3\cos^{2}\alpha-1\right),
\end{equation}
where $c$ is the speed of light in vacuum, $\Gamma_{0}$ denotes the spontaneous radiation rate of the atom, $R$ is the distance between two adjacent atoms and $\alpha$ denotes the angle of the atomic distance to the electric dipole moment (please see Fig. \ref{Fig1}). In the following, we consider that the array of interacting atoms is assumed to be in one-dimensional array with $N$ sites located at equal distances from each other. Simultaneously, we use the symbol $\eta$ to denote the interaction strength between nearest-neighbor atoms. 

From the above Hamiltonian, there are two charging mechanisms: One is to charge directly with atomic interactions; and the other is to transfer energy from the cavity field to atoms. It is worth mentioning that we attribute the indirect charging of the driving field to the second charging method. Then we further assume that atomic system is in its ground state $\ket{\textit{g},\textit{g},\textit{g}, ...,\textit{g}}$ and the cavity is in a Fock state with a specific photon number $n$, i.e., $\ket{\textit{n}}$. Therefore, the overall initial state of battery and charger is
\begin{equation}\label{eq_05}
\ket{\varphi(0)}_{s}=\ket{\underbrace{\textit{g},\textit{g},\textit{g}, ...,\textit{g}}_{N}}\otimes\ket{\textit{n}}.
\end{equation}
In our charging protocol, the QB begins to charge when $\Theta(t)$ equals one, and then the evolution of the wave function over time reads
\begin{equation}\label{eq_06}
\ket{\varphi(t)}_{s}=\mathcal{T} {\rm exp}\left[{-i\int_{0}^{t}\hat{H}_{s}(t')dt'}\right]\ket{\varphi(0)},
\end{equation}
where $\mathcal{T}$ denotes time ordering operator. Thus, the density matrix of the QB in the initial state and final state at an arbitrary time can be respectively written as
$\hat{\rho}_{s}(0)=\ket{\varphi(0)}_{s}\bra{\varphi(0)}$ and $\hat{\rho}_{s}(t)=\ket{\varphi(t)}_{s}\bra{\varphi(t)}$. In the charging process, the charger transfers its own energy to the battery, leading to the internal atoms of the battery to transition from the ground state $\ket{\textit{g}}$ to the excited state $\ket{\textit{e}}$. Therefore, the stored energy in the QB at time $t$ is given by
\begin{equation}\label{eq_07}
E_b(t)={\rm Tr}\left[\hat\rho_{b}(t)\hat{H}_{b}\right]-{\rm Tr}\left[\hat\rho_{b}(0)\hat{H}_{b}\right].
\end{equation}
Here $\hat\rho_{b}(t)={\rm Tr}_{c}[\hat\rho_{s}(t)]$ is the reduced density matrix of the QB. Remarkably, the stored energy $E_{b}(t)$ in which we are concerned does not contain the Hamiltonian $H_{\rm int}$ since the existence of atomic interactions induces either positive or negative contributions to $E_{b}(t)$, which finally results in an uneven comparison. Similarly, in order to quantify the rate of storing energy during the charging process, we define the average charging power as follows
\begin{equation}\label{eq_08}
P_{b}(t)=\frac{E_{b}(t)}{t}.
\end{equation}
When evaluating the performance of QBs in terms of stored energy and average charging power, the number of atoms in the cavity will inevitably generate energy fluctuations. To ensure the most stable charging process, we need to further analyze the superiority of the charging process by energy quantum fluctuation \cite{PhysRevB.102.245407}, which can be written as
\begin{equation}\label{eq_09}
\begin{split}
   \Delta E_{b}(t) =&\omega_{0}\Big[\sqrt{{{\rm Tr}(\hat\rho_{b}(t)\hat{H}_{b}^{2})}-[{\rm Tr}(\hat\rho_{b}(t)\hat{H}_{b})]^{2}}\\
                    &-\sqrt{{{\rm Tr}(\hat\rho_{b}(0)\hat{H}_{b}^{2})}-[{\rm Tr}(\hat\rho_{b}(0)\hat{H}_{b})]^{2}}\Big].\\
\end{split}
\end{equation}
\begin{figure}[htbp]\centering
\includegraphics[scale=0.4]{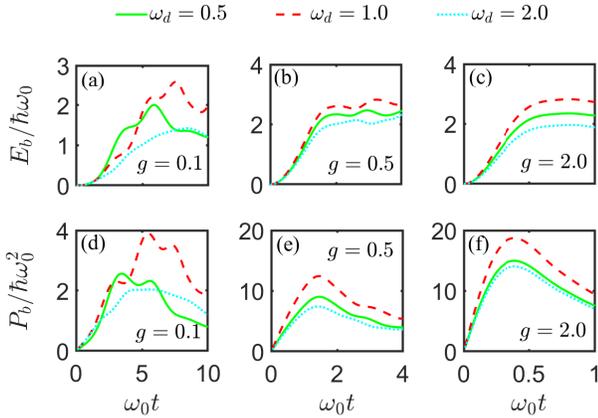}
\caption{(a)-(c) Behavior of the stored energy $E_{b}(t)$ (in units of $\hbar\omega_{0}$), and (d)–(f) behavior of the charging power $P_{b}(t)$ (in units of $\hbar\omega_{0}^{2}$) as a function of the time $t$ (in units of $\omega^{-1}_{0}$) in the case of the USC and DSC regime. Different timescales have been used to properly show the maximum position of the related-performance. The various curves in the figure represent the various frequencies of driving field and all are for the QB with the number of qubits $N$ =5.}\label{Fig2}
\end{figure}
\begin{figure}[htbp]\centering
\includegraphics[scale=0.4]{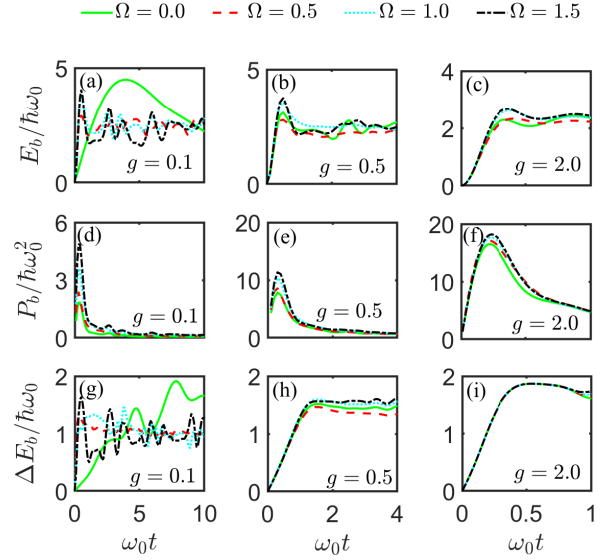}
\caption{(a)-(c) Behavior of the stored energy $E_{b}(t)$ (in units of $\hbar\omega_{0}$), (d)–(f) behavior of the charging power $P_{b}(t)$ (in units of $\hbar\omega_{0}^{2}$), and (g)-(i) behavior of the energy quantum fluctuation $\Delta E_{b}$ (in units of $\hbar\omega_{0}$) as a function of the time $t$ (in units of $\omega^{-1}_{0}$) in the case of the USC and DSC regime. Different timescales have been used to properly show the maximum position of the related-performance. The various curves in the figure represent different values of the selected $\Omega$ and all are for the QB with the number of qubits $N$ = 5.}
\label{Fig3}
\end{figure}
According to the previous studies \cite{PhysRevResearch.4.033177}, we focus on the resonant regime, as off-resonant case ($\omega_{c} \neq \omega_{0}$) gives rise to a low probability of transition, leading to less efficient energy transfer between the cavity and battery \cite{PhysRevA.104.033716}. Here and after, we assume the resonant condition, i.e., $\omega_{c} = \omega_{0} = 1$. Furthermore, to determine the optimal QB as much as possible, in Fig. \ref{Fig2} we illustrate the time evolution of the energy and charging power for different values of the driving field frequency in various coupling regimes. After comparing the results, it is evident that the performance in the other two regimes is markedly inferior to that in the resonant regime ($\omega_{d}=\omega_{0}$), in terms of both the maximum energy storage and the maximum charging power. Therefore, given the situation reported above, it is necessary to assume $\omega_{d}=\omega_{0}$ to achieve greater energy and faster charging of the QB. Besides, we also take $\Omega$ and $\eta$ in the unit of $\omega_{0}$ for convenience. For dipole-dipole interactions that decay with the cube of atomic distance, the long-range interactions have negligible effects. Hence, the atomic interaction can be ignored when separation distance is wider than four times of the distance between two nearest-neighbor atoms in the numerical simulation.
\begin{figure}[htbp]\centering
\includegraphics[scale=0.4]{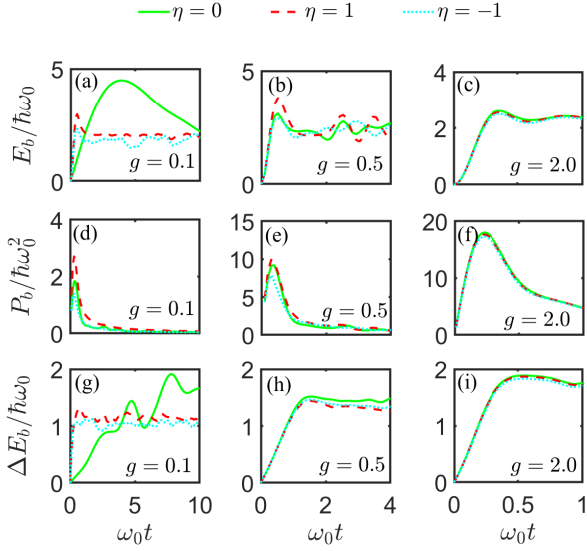}
\caption{(a)-(c) Behavior of the stored energy $E_{b}(t)$ (in units of $\hbar\omega_{0}$), (d)–(f) behavior of the charging power $P_{b}(t)$(in units of $\hbar\omega_{0}^{2}$), and (g)-(i) behavior of the energy quantum fluctuation $\Delta E_{b}$ (in units of $\hbar\omega_{0}$) as a function of the time $t$ (in units of $\omega_{0}^{-1}$) in the case of the USC and DSC regime. Different timescales have been used to properly show the maximum position of the related-performance. The various curves in the figure represent the various strengths of atomic interaction and all are for the QB with the number of qubits $N$ = 5.}
\label{Fig4}
\end{figure}

Furthermore, it should be emphasized that our study of the QB is conducted within a closed system. It may inevitably interact with the external environment, which leads to the loss of photons in the cavity and the relaxation of the atomic energy in the QB \cite{PhysRevE.104.044116}. Therefore, in the following text, we will ensure that the evolution time $t$ is significantly shorter than the dissipation timescales $t_{\gamma}$ and $t_{r}$, so that we can safely disregard the effect of dissipation. Based on the known experimental level of operable qubits, this condition is typically fulfilled in the most sophisticated circuit quantum electrodynamics devices \cite{PhysRevLett.120.227701,PhysRevB.94.205421,PhysRevX.7.011030,RevModPhys.93.025005}. Not only that, since the general condition $t_{r} > t_{\gamma}$, the stability of the cavity becomes crucial in experiments \cite{PhysRevA.76.042319,PhysRevA.102.060201}.

Notably, when the system enters the USC and DSC regimes, we need to investigate the performance of the QB in both cases. To ensure consistency with the approach used for the Dicke QB, we remark the existence of the counter-rotating term in the Hamiltonian of the system, and the total excitation number is not conserved according to the Hamiltonian of Eq. (\ref{eq_02}). Thus, in principle, we can choose any large number of photons. However, in practical finite-size numerical simulation, we need to introduce a maximum photon number $N_{\rm ph}$ to ensure that the related-performance of the QB is calculated accurately \cite{PhysRevLett.120.117702}. We have extensively examined the numerical convergence of results (energy, charging power, and energy quantum fluctuation) and demonstrated the error below $10^{-5}$ by selecting a relatively large photon number $N_{\rm ph}=4N$ \cite{PhysRevB.102.245407,PhysRevB.105.115405}.
\section{The influence of external driving field and dipole-dipole interaction on the QB at different coupling strengths}
\label{SecIII}
In this section, we will separately analyze the influence of the driving field amplitude, the atomic interaction, and the combined actions on the performance of the QB in the USC ($0.1\omega_{0} \leq g \lesssim \omega_{0}$), and DSC ($g \geq \omega_{0}$) regimes. We expect higher energy storage and further enhancement of the average charging power compared to the Dicke QB.
\subsection{Influence of external driving field on the performance}
We now consider that the distance between atoms is sufficiently large, and dipole-dipole interaction can be ignored. Acting external driving field on the system, the total Hamiltonian during the charging process reads
\begin{equation}\label{eq_10}
{\hat{H}_{s}}=\omega_{0}\hat{J}_{z}+\omega_{c}\hat{a}^{\dagger}\hat{a}+2g(\hat{a}^{\dagger}+\hat{a})\hat{J}_{x}+\Omega\cos(\omega_{d}t)(\hat{a}^{\dagger}+\hat{a}).
\end{equation}
In Fig. \ref{Fig3}, we show the influence of variations in the driving field strength on the stored energy, charging power, and energy quantum fluctuation in the USC and DSC regimes. When the cavity-atom coupling is not very strong, regardless of whether the external driving field strength is large or small throughout the whole charging process, it can play a positive role in charging power based on Eq. (\ref{eq_10}) in comparison to Dicke QB (green line). However, in the USC regime, as the driving field amplitude increases, the stored energy and average charging power also change significantly accordingly, but the influence of this enhancement becomes smaller compared with that of the weak USC case. In particular, when the amplitude raises at a fixed value, the maximum charging power of the QB does not uniformly increase. More interestingly, in the USC regime, the upgrade of the coupling strength allows the QB to achieve the maximum stored energy more quickly than in the weak USC regime.

Next, we discuss the coupling strength in the DSC regime, specifically up to $g \geq \omega_{0}$. However, the changes in the effective amplitude will not strongly impact the performance of the QB in the DSC regime. Despite this, the corresponding performance of the QB has improved as a result of the increased amplitude. Such situation can be explained as follows: when the coupling between cavity and atoms is weak, the driving term dominates the dynamics; nevertheless, as the coupling strength gradually increases, the indirect charging dominated by the driving field has no advantages over the cavity-atom charging. Thus, the QB does not change with the driving field amplitude in the DSC regime. On the other hand, compared to the weak USC regime, the QB also generates higher energy quantum fluctuation during charging. Although the quantum fluctuation is inevitable and limited, this phenomenon should be caused by the interaction between atoms and the cavity field, which is closely related to the QB not being fully charged.
\begin{figure}[htbp]\centering
\includegraphics[scale=0.4]{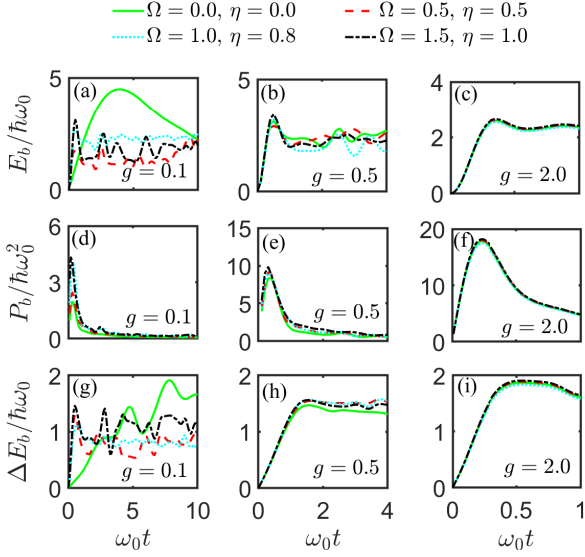}
\caption{(a)-(c) Behavior of the stored energy $E_{b}(t)$ (in units of $\hbar\omega_{0}$), (d)–(f) behavior of the charging power $P_{b}(t)$ (in units of $\hbar\omega_{0}^{2}$), and (g)-(i) behavior of the energy quantum fluctuation $\Delta E_{b}$ (in units of $\hbar\omega_{0}$) as a function of the time $t$ (in units of $\omega^{-1}_{0}$) in the case of the USC and DSC regime. Different timescales have been used to properly show the maximum position of the related-performance. The various curves in the figure represent different values of the selected $\Omega$, $\eta$ and all are for the QB with the number of qubits $N$ = 5.}
\label{Fig5}
\end{figure}
\subsection{Influence of dipole-dipole interaction on the performance}
We now assess the performance of the QB for the Dicke model with dipole-dipole interaction between atoms. The total Hamiltonian based on Eq. (\ref{eq_02}) during the charging process is given as
\begin{equation}\label{eq_11}
\hat{H_{s}}=\omega_{0}\hat{J}_{z}+\omega_{c}\hat{a}^{\dagger}\hat{a}+2g(\hat{a}^{\dagger}+\hat{a})\hat{J}_{x}+\sum_{i\neq j}^{N}\eta_{ij}(\hat{\sigma}_{i}^{-}\hat\sigma_{j}^{+}+\hat{\sigma}_{j}^{-}\hat\sigma_{i}^{+}).
\end{equation}
In Fig. \ref{Fig4}, we show the influence of atomic interaction on stored energy, charging power, and energy quantum fluctuation in the USC and DSC regimes. The red and blue lines indicate the repulsive ($\eta > 0$) and attractive ($\eta < 0$) interactions, respectively. For comparison, the performance of Dicke QB is shown with solid green line. When the QB is in the weak USC regime, the interactions always have disadvantages on the maximum stored energy, which can be attributed to a portion of the energy is stored into the interaction between atoms. So, we can identify that $\eta$ has the disadvantage of energy storage over Dicke QB. However, for the charging power, different atomic interactions have different effects, in which the attractive interaction decreases and repulsive interaction increases the maximum charging power. It can be explained that the repulsive interaction increases the effective atom-field interaction; while the attractive interaction suppresses the effective interaction \cite{PhysRevA.78.023634}. On the other hand, quantum correlations during charging can enhance the rate at which energy can be stored, usually denoted as a quantum speed up \cite{PhysRevResearch.2.023113}. Therefore, different $\eta$ lead to varying degrees of increasing or decreasing the charging power.

Nevertheless, the results show that the influence of the atomic interaction on the performance is not significant in the DSC regime. It is shown that the advantage of the atomic interaction on the QB has become smaller and smaller, indicating competition between atom-atom and atom-cavity interactions \cite{PhysRevA.78.023634,PhysRevB.105.115405}. The interaction between atoms plays a dominant charging mechanism comparing to the atom-cavity interaction in the USC regime, making it a crucial factor influencing the QB. However, in the DSC regime, the interaction between atoms and the cavity field is strong enough to counteract the benefits generated by the atomic interaction, resulting in minimal impact on performance as $\eta$ changes. Correspondingly, we observe that the maximum stored energy exhibits a critical behavior, that is, the QB system exists at the critical point where the maximum energy storage changes significantly. We already know that spin-spin interactions can produce ground-state quantum phase transition (QPT) from Refs. \cite{PhysRevA.103.033715,PhysRevA.84.053610,PhysRevE.67.066203,PhysRevA.78.023634,PhysRevLett.90.044101}. Since we assume that all the qubits in the initial state are oriented downward, the original symmetry is broken which affects the ground state properties during charging. The QPT also influences the energy charging and release of the QB (More details are presented in Appendix \ref{appendix1}).
\begin{figure}[htbp]\centering
\includegraphics[scale=0.6]{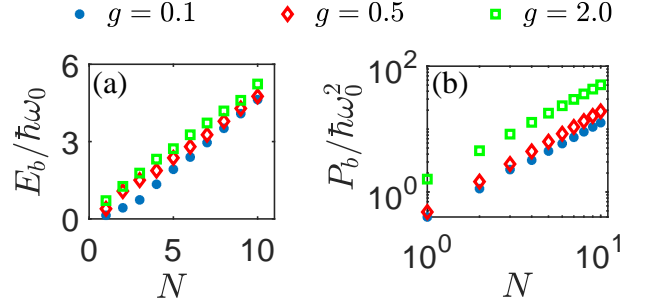}
\caption{(a) Maximum stored energy $E_{\rm max}$ (in unit of $\hbar\omega_{0}$), and (b) maximum charging power $P_{\rm max}$ (in unit of $\hbar\omega^{2}_{0}$) of QB with the number of qubits $N$ for different coupling regimes $g$. The other parameters are valued as $\Omega = 1.0$, $\eta = 0.8$.}
\label{Fig6}
\end{figure}
\subsection{Influence of dipole-dipole interaction and external driving field on the performance}
\begin{table*}[htbp]
\centering
\caption{The scaling exponent $\alpha$ of the maximum charging power for different values of the cavity-spin coupling strength $g$, the effective amplitude of external driving field $\Omega$ and the interaction strength $\eta$.}
\label{tab1}
\setlength{\tabcolsep}{1.4mm}
\begin{tabular}{ccccccccccccc}
\hline
\hline
 &  &$\eta=-1.0$&  &  &$\eta=-0.5$&  &  &$\eta=+0.5$& & &$\eta=+1.0$\\
 \cmidrule(lr){2-4}\cmidrule(lr){5-7}\cmidrule(lr){8-10}\cmidrule(lr){11-13}
               & $g=0.1$ & $g=0.5$ & $g=2.0$ & $g=0.1$ & $g=0.5$ & $g=2.0$& $g=0.1$ & $g=0.5$ & $g=2.0$& $g=0.1$ & $g=0.5$ & $g=2.0$\\
\hline
  $\Omega=0.1$ & 1.59 & 1.20 & 1.03 & 1.55 & 1.17 & 1.02 & 1.57 & 1.19 & 1.02 & 1.60 & 1.23 & 1.05\\            
  $\Omega=0.5$ & 1.47 & 1.43 & 1.33 & 1.53 & 1.42 & 1.32 & 1.55 & 1.44 & 1.34 & 1.49 & 1.45 & 1.34\\          
  $\Omega=1.0$ & 1.45 & 1.45 & 1.40 & 1.51 & 1.43 & 1.38 & 1.53 & 1.45 & 1.36 & 1.47 & 1.46 & 1.42\\
  $\Omega=2.0$ & 1.40 & 1.47 & 1.47 & 1.43 & 1.45 & 1.45 & 1.45 & 1.48 & 1.44 & 1.37 & 1.48 & 1.46\\               
\hline
\hline\end{tabular}\end{table*}
In this section, we investigate the performance of the QB for the Dicke model with an external driving field and the atomic interaction. Accordingly, the Hamiltonian during the charging process reads
\begin{equation}\label{eq_12}
\begin{split}
\hat{H_{s}}=&\omega_{0}\hat{J}_{z}+\omega_{c}\hat{a}^{\dagger}\hat{a}+2g(\hat{a}^{\dagger}+\hat{a})\hat{J}_{x}+\Omega\cos(\omega_{d}t)(\hat{a}^{\dagger}+\hat{a})\\
&+\sum_{i\neq j}^{N}\eta_{ij}(\hat{\sigma}_{i}^{-}\hat\sigma_{j}^{+}+\hat{\sigma}_{j}^{-}\hat\sigma_{i}^{+}).
\end{split}
\end{equation}
In Fig. \ref{Fig5}, we show the behavior of the stored energy $E_{b}$, average charging power $P_{b}$, and energy quantum fluctuation $\Delta E_{b}$ for different values of $\Omega$ and $\eta$ based on Eq. (\ref{eq_12}) in the USC and DSC regimes during the charging process. For comparison purposes, the case without driving field and atomic interaction is also depicted by the solid green line. In the weak USC regime, the maximum energy storage and the maximum charging power become weakened compared to the single driving field $\Omega$ action previously shown in Fig. \ref{Fig3}, while the charging process becomes more stable and then the time for fully charging is indeed reduced accordingly; however, the QB exceeds the single atomic interaction $\eta$ previously shown in Fig. \ref{Fig4} in terms of the maximum energy storage and the maximum charging power, but the time to complete charging is extended. With the increase in the coupling strength $g$, the combined actions of the two also affect the performance of the QB to varying degrees. Similarly, in the DSC regime, changing the effective amplitude of the driving field $\Omega$ and the atomic interaction strength $\eta$ has minimal impact on the related-performance of the QB. It can also be explained that the cavity-atom charging outperforms the charging dominated by the driving field (atomic interaction) with the gradual increase in coupling strength. After comparison, we find the improvement of the charging performance is the most obvious in the weak USC regime. Of the four curves listed, when $\Omega = 1.0, \eta = 0.8$, the performance of the QB has a significant advantage, whereas the other curves are not comparable, indicating the need for appropriate adjustment of $\Omega$ and $\eta$ to achieve the optimal performance. Therefore, combined with practical experimental implementations, it is essential to rationally design the QB in the weak USC regime to achieve greater energy storage and faster charging rate. 
\section{Enhancement of collective charging}\label{SecIV}
In this section, we further discuss how the performance of the QB depends on the number of qubits in the presence of both driving field and atomic interaction. In Sec. \ref{SecIII}, the selected parameters, namely $\Omega=1.0$ and $\eta=0.8$, have a significant impact on all aspects of the performance in the weak USC regime. So we take these values as the basic parameters to study how the number of qubits $N$ affects the charging process. In Fig. \ref{Fig6}, we illustrate the dependence of the energy storage and charging power on the number of qubits. Therefore, we can easily observe that the maximum energy storage of the QB increases with the number of qubits $N$. At the same time, we can fit the data to find that the maximum energy storage of the current QB is proportional to the number of qubits ($E_{\rm max}\varpropto N$), which closely resembles the conclusion previously described in the literature \cite{PhysRevA.106.032212,PhysRevLett.120.117702}.

The trend of the charging power follows that of the stored energy in Fig. \ref{Fig6}. However, relying solely on the tendency of the graph line is far from insufficient, and we need to introduce another parameter to compare the performance of the QB. Since the charging power of the QB is limited by the charging Hamiltonian added to the battery in the external quenching charging protocol, we can apply the quantum advantage $\alpha$ proposed in Refs. \cite{PhysRevLett.120.117702,PhysRevB.105.115405}. The scaling index $\alpha$ essentially reflects the property of collective charging of the QB. We already know that the quantum advantage $\alpha$ is an upper bound on the quadratic function of the maximum participation number of the collective charging \cite{PhysRevLett.118.150601, PhysRevLett.128.140501}. Furthermore, in the Dicke QB, the maximum charging power $P_{\rm max}$ is proportional to $N^{\frac{3}{2}}$, suggesting that the maximum charging power of our QB may also exhibit a scaling relationship with the number of qubits $N$. We assume that the following form can be used for the maximum charging power ($P_{\rm max}\varpropto\beta N^{\alpha}$). By properly adjusting the parameters, we can bring the $\alpha$ to 1.60 in the range considered in Tab. \ref{tab1} (More details are reported in Appendix \ref{appendix2}). This result is consistent with the conclusion previously reported in the reference \cite{PhysRevB.105.115405}. On the other hand, setting $g=0.1$, $\Omega=0.1$, $\eta=1$ corresponds to a smaller driving field and larger atomic interaction, which could effectively reduce energy oscillation to achieve a relatively stable charging process in the weak USC regime. So we fully predict that the QB can store more energy and achieve higher charging efficiency.
\section{Conclusion}\label{SecV}
We consider a QB described by introducing the external driving field and dipole-dipole interaction between atoms in the Dicke model. When only the external driving field exists, the QB always shows better stored energy and charging power as the effective amplitude increases; while considering only dipole-dipole interaction between atoms, the performance of the QB is influenced by the atomic interaction in a different way. However, when both external driving field and atomic interaction exist, we choose some parameters and find the increase advantage of $\eta$ and $\Omega$ is not the same, which requires to properly design the parameters to obtain the optimized QB. Furthermore, we have observed that the maximum stored energy exhibits a critical behavior. The critical atomic interactions and coupling strengths have been analyzed by QPT. We also further examine the impact of the number of qubits $N$ on the performance of the QB. Upon comparison with the traditional Dicke QB, we obtain a quantum advantage of the maximum charging power, approximately satisfying $P_{max}=\beta N^{\alpha}$ and find the quantum advantage of the maximum charging power of our QB can reach 1.60 by adjusting the corresponding parameters appropriately. So we can fully believe in the implementation of greater energy storage, faster charging rate, and a more stable process for QB. It is hoped that our proposed scheme can provide valuable references for the future development of the QB.
\section*{Acknowledgments}
Chunfeng Wu is supported by the National Research Foundation, Singapore and A*STAR under its Quantum Engineering Programme (NRF2021-QEP2-02-P03).
\appendix
\renewcommand{\appendixname}{Appendix}
\section{Critical behavior of maximum energy storage}
\label{appendix1}
The QPT is the phenomenon of phase transition occurring at zero temperature due to quantum fluctuations, contrasting thermal phase transitions which are caused by thermal disturbances. It represents a sudden burst of ground state properties in many-body quantum systems with external parameters. The QPT exists widely in magnetic materials, ferroelectric materials, metal-insulator transition systems, and is an important foundation for a deeper understanding of condensed matter physics.
\begin{figure}[htbp]\centering
\includegraphics[scale=0.42]{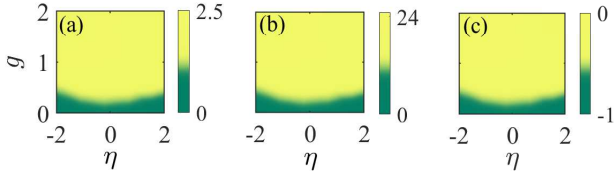}
\caption{(a)-(c) Behavior of maximum stored energy $E_{\rm max}$ (in units of $\hbar\omega_{0}$), maximum charging power $P_{\rm max}$ (in units of $\hbar\omega^{2}_{0}$) and the phase diagram $\langle \hat{J}_{z} \rangle/(N/2)$ as the function of the atomic interaction $\eta$ and the coupling strength $g$. All plots correspond to the case of $N = 5$.}
\label{Fig7}
\end{figure}
For our QB system, atom-atom interactions can significantly change the maximum stored energy of the QB in the USC regime. However, in the DSC regime, such influence is negligible. In particular, the maximum stored energy exhibits a critical behavior in the USC regime (There is a critical regime in the system, and the maximum stored energy of the QB changes significantly around the critical regime). Since in our QB charging scheme, the energy storage of the system mainly depends on the state of all qubits in the cavity. Therefore, we discuss the QPT in our analysis. To clarify the behavior, we treat the QB as only two independent parameters, namely the atomic interaction $\eta$ and the coupling strength between atoms and the cavity $g$.

In Fig. \ref{Fig7}, since the maximum energy storage $E_{\rm max}$ and the maximum charging power $P_{\rm max}$ are measured during the evolution process, it is not clear whether they can illustrate the QPT. To better illustrate the QPT behavior, we introduce an ordered parameter, namely the $z$ component of the average magnetization $\hat{J}_{z}$. Therefore, the curve from the ferromagnetic phase ($\langle \hat{J}_{z} \rangle/(N/2)= -1$) to the Mott phase ($\langle \hat{J}_{z} \rangle/(N/2)= 0$) is depicted in Fig. \ref{Fig7}. This means that when $\langle \hat{J}_{z} \rangle/(N/2)= -1$, the system is well-ordered where all the qubits are in the spin-down state of the ferromagnetic phase. However, when the parameter changes so that $\langle \hat{J}_{z} \rangle/(N/2)= 0$, the system begins to deviate from the ferromagnetic phase to reach the Mott phase. At this point, the order of the system is broken and presents the antiferromagnetic phase. We already know that $\hat{J}_{z}$ is the layout difference operator between the excited state and the ground state, and that its value is related to the number of atoms in each state. The Mott phase represents an equal number of atoms in both states; while the ferromagnetic phase means that all atoms are in the ground state. As the ground state we set is all spin-down, the QB in the ferromagnetic phase is detrimental for storing energy. Correspondingly, the maximum stored energy changes significantly at the critical point of the QPT.\\
\section{Collective enhancement of the charging power}
\label{appendix2}
In order to further discuss how the charging process of the QB depends on the number of qubits when the charger remains fixed, we now analyze the scaling of maximum charging power $P_{\rm max}$ based on Eq. (\ref{eq_08}) as a function of the number of the qubits $N$. In Fig. \ref{Fig6}, it can be found that the maximum stored energy and the maximum charging power have a clear correlation with $N$ as it increases. We recall that in Ref. \cite{PhysRevLett.120.117702}, it has been shown that the energy scales extensively with $N$, while the power shows a superlinear behavior with $N$. The behavior of the maximum charging power reflects the acceleration effect of the collective scheme and shows the quantum advantage. Although it has been pointed out in the literature \cite{PhysRevLett.118.150601} that the increased quantum advantage associated with the increasing number of cells in the cavity is impossible to be achieved at any physically reasonable Hamiltonian (The Hamiltonian with the most $N$-body interactions). However, with increased interactions between QB during collective charging, the quantum advantage can be achieved. We already know the maximum charging power of the Dicke QB ($P \propto N^{\frac{3}{2}}$), so naturally assume that the maximum charging power takes the following form
\begin{equation}\label{C1}
P_{\rm max}\varpropto\beta N^{\alpha},
\end{equation}
By taking the logarithm, we use linear fitting to obtain the scaling exponent $\alpha$
\begin{equation}\label{C2}
\log(P_{\rm max})=\alpha\log(N)+\log(\beta).
\end{equation}
 If $\alpha$ has greatly surpassed $1.5$ of the corresponding Dicke QB, further demonstrating the superiority of our QB.
\bibliography{reference}
\end{document}